# Quantifying Stellar Mass Loss with High Angular Resolution Imaging

Contact: Stephen Ridgway
ridgway@noao.edu

Coauthors and Contributors

Jason Aufdenberg (Embry-Riddle Aeronautical U),
Michelle Creech-Eakman (New Mexico Tech),
Nicholas Elias (U Heidelberg),
Steve Howell (NOAO),
Don Hutter (USNO)
Margarita Karovska (Harvard-Smithsonian CfA),
Sam Ragland (Keck Observatory),
Ed Wishnow (U California Berkeley),
Ming Zhao (U Michigan)

**Introduction.** Mass is constantly being recycled in the universe. One of the most powerful recycling paths is via stellar mass-loss. All stars exhibit mass loss with rates ranging from ~$10^{-14}$ to $10^{-4}$ $M_\odot$ yr$^{-1}$, depending on spectral type, luminosity class, rotation rate, companion proximity, and evolutionary stage. The first generation of stars consisted mostly of hydrogen and helium. These shed material – via massive winds, planetary nebulae and supernova explosions – seeding the interstellar medium with heavier elements. Subsequent generations of stars incorporated this material, changing how stars burn and providing material for planet formation.

An understanding of mass loss is critical for modeling individual stars as well as answering larger astrophysical questions. Understanding mass loss is essential for following the evolution of single stars, binaries, star clusters, and galaxies. Mass loss is one of our weakest areas in the modeling of fundamental stellar processes. In large part this is owing to lack of confrontation with detailed observations of stellar photospheres and the mass-loss process. High resolution optical imagery with telescope arrays is beginning to provide these data and, combined with spectroscopy and broad infrared and sub-mm coverage, supporting more sophisticated models on fast computers and promising a new era in mass-loss studies.

**Why is Stellar Mass Loss Important?**

Stellar mass loss is key to understanding the evolution of the universe from the earliest cosmological times to the current epoch, and of planet formation and the formation of life itself. The mass loss in the original massive stars was the crucial "ingredient" in shaping the early galaxy formation and the current universe. Mass loss determines pre-main sequence evolution (Weidenschilling, EarthMoon&Planets 19, 279 (2004)), time on the main sequence (Mitalas & Falk, MNRAS 210, 641 (1984)), the horizontal branch morphology (Dotter, arXiv:0809.2603v1 (2009)), the formation and characteristics of Planetary Nebulae (Harpaz, ApJ, 498,293, (1998)), impacts the pre-explosion characteristic of SNII (Taylor, "The Stars", Cambridge (1994)), and the remnant shells with which they interact. Mass loss from massive stars must be included to understand the evolutionary connections between Be stars, W-R stars and Hypergiants (Lamers, ASP Conf 388, 443 (2008)), as well as other massive, hot stars.

Interacting binaries exhibit both mass loss and mass transfer, which means that their component stars have outer atmospheres which may be inconsistent with their interiors for a given age. Mass loss may be more important for the evolution and stability of star clusters than dynamics alone (van Loon & McDonald, ASP Conf 388, 405 (2008)). Mass loss from red giant stars is a major source of key elements, including C isotope abundances in the galactic disk, and N and O abundances in different types of galaxies (Tosi, ASP Conf 378, 353 (2007)). The metallicity dependence of mass loss is vital to understanding diverse topics such as the precursors to GRB's, and early galactic chemical enrichment (Maeder, ASP conf 353, 415 (2006)).

**Current Limitations and What Do We Want to Learn About Mass Loss?**

*The last few days we heard many talks about mass loss rates, which together present a nice state-of-the-field review. What is my impression? The topic is even more uncertain than it was before!*

H. Lamers, 2008, ASPC 388, 443

Our measurements of mass loss are traditionally based on emission and scattering from circumstellar shells and on spectral line profiles, in each case representing an uncertain averaging over the star's history and circumstellar volume. In even simple scenarios, such information cannot uniquely determine mass-loss parameters. The stellar regions where grain formation take place are uncertain, probably lying within a few stellar radii of the surface, or possibly in the atmosphere itself. The (a)symmetry of mass loss is increasingly noted, but association to a particular cause is mostly speculative. The difficulties in extracting physical parameters from photometric and spectroscopic measurements alone are at the root of the frustration expressed by Henny Lamers after working on mass loss in hot stars for more than 40 years.

Understanding mass loss in cool stars has long been impeded by lack of computer speed and insufficient observational verification. The field is now poised to move forward on both accounts, i.e. advances in theoretical modeling and interferometric high-angular resolution observations:

- It is becoming feasible to connect the interior pulsation (Cox, 2002, ASP Conf 259, 21) and convective physics (Hervig et al. 2007, ASPC 378, 43) with the 3D surface dynamics and non-equilibrium processes, including radiative transfer (Hoefner, 2008, A&A 491L, 1) and grain growth (Sedlmayr et al. 2004, EAS 11, 5). The goal of such coupled and self-consistent modeling will be to explain observed patterns in mass loss (Willson, 2007, ASPC 378, 211), to confirm, adapt and extend empirical models to a broader range of application, and eventually to account for realistic mass-loss activity in the evolution of single stars, binary systems, and the evolution of galactic abundances and stellar populations.

- The potential of interferometric observations of mass loss - Optical interferometry is one of the most important tools for investigating mass-loss available today. From UV through optical and mid-infrared and out to radio wavelengths, multiple excited emission lines and molecular band-heads are available to probe the entire environment of nearly any type of mass-losing star at resolutions of ≤ milliarcsecond (often less than 1 stellar radius). The spatial locations of the atomic lines, molecular bands and dust features can be used as direct input and challenges to existing theoretical models of the most fundamental process in all stars after nucleosynthesis. When coupled with spectro-interferometric measures, or directly with other techniques such as spectroscopy, tomography, reverberation mapping, polarization and high resolution VLBI radio images of continuum emission or masers, optical interferometry allows astronomers to create a full and contemporaneous map of the circumstellar environment, including wind structures and episodic histories. No longer do stars need to appear as static entities requiring many years and/or statistical samples to create a complete picture. Questions regarding asymmetries and their onset, locations of the bases of winds, the onset of dust formation, and the episodic nature of stellar phenomena, can now be studied directly as they are occurring rather than inferred from averaged signatures. Many of the phenomena linked to mass-loss, e.g. pulsation, spots, convection and magnetic fields, are indeed the most compelling when examined in the spatially resolved time-dependent domain. This field of study is readily opened with the use of the sub-milliarcsecond spatial resolutions employed now routinely by optical/IR interferometers, which will only continue to increase in diagnostic power over the next decade as imaging techniques grow in sophistication and use.

The following pages will address in more detail several specific mass loss topics that can be addressed in the upcoming decade.

**Mass Loss from Red (Super)giants:** features low wind speeds and the lack of UV continuum photons to couple to the atomic resonance transitions – these winds are neither energetic nor radiatively driven. Indeed, no known single/combination of mechanisms can explain the observed wind characteristics witnessed today (c.f Harper, 1996, ASPC 109, 481). Multiple techniques are currently used to characterize the mass loss including: a) observing optically thick free-free emission in the radio, b) examining UV Mg II absorption lines to get wind terminal velocities, c) deriving wind temperatures from ionized species at less than 10,000K, d) inferring wind structures from episodic behavior and variability in these lines, and e) characterizing long-term behavior from infrared maps and observations. Some success has been seen examining widely separated symbiotic systems like EG And (Espey & Crowley, 2007, ASP, "RS Oph meeting", in press) where the wind from the red giant is illuminated by the UV continuum of the distant white dwarf companion. Espey's results, in particular, show that the wind acceleration starts in an isothermal medium as close as 2 stellar radii to the star, and is clumpy with velocities in the 3-5 km/s regime. One tantalizing explanation offered for the observations is the presence of Alfven waves, but only if they can be sufficiently damped in order to match observed terminal velocities (c.f. Hartmann & MacGregor, 1980, ApJ, 242, 260). On scales of several arcseconds, there is mounting evidence for spatially resolved IR CO lines around a handful of red supergiants, in distributions which are spherically symmetric with interspersed asymmetric clumps, arising from presumably long-term episodic mass loss (Smith et al., 2009, AJ, in press).

Contrast these results using traditional techniques to the optical/IR interferometric observations of red giants and supergiants taken over the last decade, which build a consistent picture of their environments. Optical interferometry has already demonstrated evidence for wavelength dependent limb darkening and a ~2000K layer of water surrounding some supergiant stars at less than one stellar radius (Perrin et al. 2004, A&A, 418, 675) and multiple time-variable spots on Betelgeuse presumed to be evidence of convection, along with evidence for formation of a dusty layer (Willson et al. 2007, MNRAS, 291, 819). In the mid-infrared, evidence for silane, hydroxyl, water and silicate dusts in these stars are challenging theories as to evolutionary tracks, mass-loss rates and progenitor masses themselves (Ohnaka et al. 2008, A&A, 484, 371; Monnier et al. 1999, AAS, 195, 4511). The observed large mid-infrared opacity of α Ori based on interferometric and spectroscopic measurements is interpreted as the presence of hot amorphous alumina in the extended atmosphere of this supergiant (Verhoelst et al. 2006, A&A 447, 311). Subsequently, molecular and dust composition at the inner regions of the extended atmosphere of this supergiant have been established from these measurements (Perrin et al. 2007, A&A 474, 599). The observed optical depths compared to the best-fitting current model opacity profile are shown in Figure 1.

Mass-loss rates and the composition of the dust shells have been estimated from interferometric observations at 11 μm (Tevousjan et al. 2004, ApJ 611, 466). Various molecular constituents have been detected at specific distances from red supergiants (Quirrenbach et al. 1993; Danchi et al. 1994), (Perrin et al. 2004) providing information about the density and temperature profiles of their massive winds which carry processed material formerly from their interiors. And most recently, the circumstellar environment of the red supergiant WHO G64 in the LMC has been resolved using mid-infrared long baseline interferometry (Ohnaka et al. 2008, A&A 484, 371) - the first resolved stellar observations in an extragalactic system. The interferometric measurements in conjunction with the spectral energy distribution and radiative transfer modeling indicate that WHO G64 may be experiencing violent, unstable mass loss. A spherical model is not satisfactory, and alternate models suggest mass loss strongest in a toroidal belt viewed approximately along the axis. The picture painted by optical interferometry is of consistent, multi-species mass loss extending from very near the star to hundreds of AU.

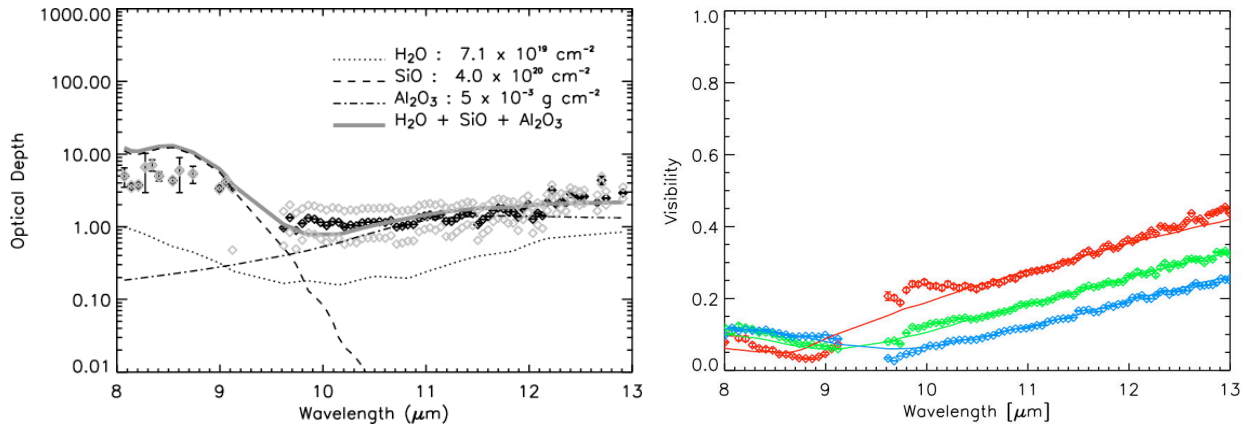

**Figure 1**. Mid-IR (MIDI) measurements of the atmosphere and envelope of α Ori. (left) Observed optical depths (black diamonds - optimum temperatures; grey diamonds - upper and lower limits on temperatures) compared to the best fitting model opacity profile (thick grey solid curve). (right) Corresponding observed and model visibility functions. (Perrin et al, 2007, A&A 474, 599). Systematic errors near 9-10 μm may reflect imperfect representation of the silicate band.

**Mass loss from Asymptotic Giant Branch Stars in the LPV Phase.** Late-type stars experience vigorous mass loss as they pass through the asymptotic giant branch (AGB) phase. Long baseline interferometric observations provide spatial resolutions sufficient to probe the complex atmospheric structure of these stars during this period of instability and rapid evolution which extends from the base of the atmosphere, where the optical depth become large (even in continuum wavelengths), to the top of the pulsing atmosphere near the radius of SiO maser emission. Such observational studies are in their infancy, but already have distinguished for the first time the true stellar surface from the extended atmosphere. In the case of Mira type stars, they have settled the question of the primary pulsational mode (in the fundamental frequency), and the typical $T_{eff}$ to be associated with interior models.

A substantial fraction of AGB stars have variable surface irregularities likely arising in heterogeneous atmospheres (Figure 2). Near-IR imaging has extended observation of asymmetry to the tenuous upper atmosphere. Ragland et al. (2008, ApJ 679, 746) probed the high $H_2O$ layer of the Mira R Aqr in the continuum and two molecular bands. The reconstructed images at three wavelengths suggest that the water molecular shell is highly clumpy (Figure 3).

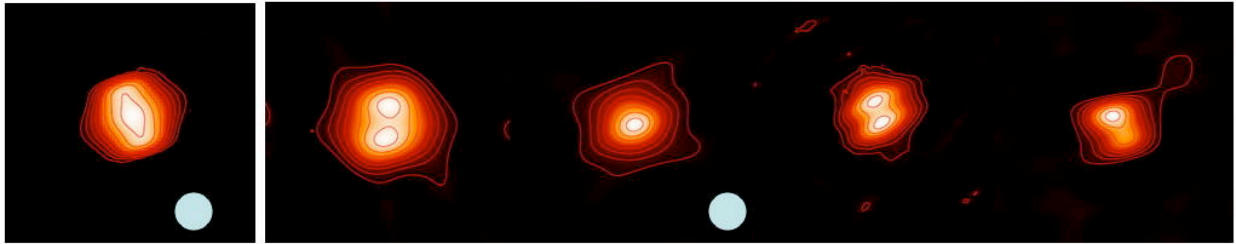

**Figure 2.** (Left) Image of Arcturus, showing the irregular PSF due to incomplete UV-plane coverage. (Right) Images of the AGB star χ Cygni, with similar UV-sampling, at phases 0.24, 0.67, 0.76, and 0.91, showing irregular structure (Lacour 2009, in preparation). The circles represent 10 milliarcseconds, approximately the angular resolution of the observations. Observing wavelength 1.6 μm (near the stellar atmospheric opacity minimum). Data are from the CFA Infrared Optical Telescope Array (IOTA) on Mt Hopkins.

With mid-IR interferometric imaging, asymmetries in Mira and supergiant stars have been studied in these layers and higher, most likely connecting to the beginnings of the dust envelopes and the wind acceleration region (Weiner et al. 2006, ApJ 636, 1067).

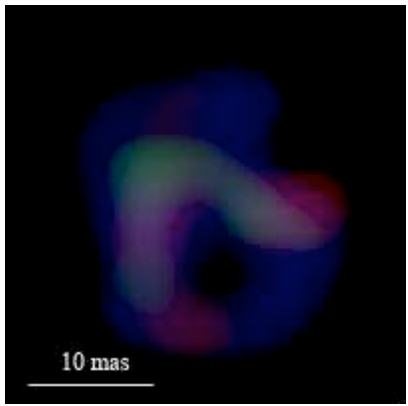

**Figure 3.** A composite reconstructed image of outer envelope of the AGB Mira star R Aqr, using long baseline interferometry. The blue, green and red colors represent 1.51, 1.64 and 1.78 μm respectively. The 1.51 μm image shows a bipolar feature roughly parallel to the large scale jets of R Aqr. Data are from the CFA Infrared Optical Telescope Array (IOTA) on Mt Hopkins, recently decommissioned.

**Blue Supergiants** are the brightest stars at visual wavelengths and among the brightest single stars visible in galaxies. Interest in the radiation-driven mass loss from blue supergiants has been motivated by their status as type II supernova progenitors and their potential as independent distance indicators via the wind momentum-luminosity relationship (Kudritzki et al. 1999, A&A

350, 970). The wind momentum, the product of the mass-loss rate and the wind terminal velocity, has been shown to correlate with the stellar luminosity in late B- and early A-type supergiants. The luminosity, predominantly ultraviolet flux, transfers momentum to the wind by radiation pressure via ultraviolet resonance lines. The spectroscopic signature of this mass-loss is the familiar P-Cygni spectral line profile.

Very-high spatial resolution observations of blue supergiants are just beginning, yet already they are shaping our understanding of the mass-loss process. The angular diameters of Deneb (A2 Ia) and Rigel (B8 Ia) have been measured (Aufdenberg et al. 2008, in *The Power of Optical/IR Interferometry,* ESO Symposium, p.71) and for Deneb the data indicate this star is not spherical, implying rotational distortion. All stellar modeling to exploit the wind momentum-luminosity relationship has thus far assumed spherical models.

Resolving the center-to-limb intensity profile for stars with compact, essentially plane-parallel atmospheres, yields information on the temperature structure of atmosphere. For blue supergiants, with highly extended atmospheres, precision center-to-limb intensity profiles will provide a mass-loss rate independent from spectroscopic and radio continuum diagnostics. Furthermore, long-baseline interferometry at high spectral resolution is just now coming on-line. Observations at a spectral resolution of R=30,000 and sub-milliarcsecond spatial resolution will test the reliability of the spatially averaged P-Cygni line profiles as mass-loss rate diagnostics and will characterize the asymmetric nature of the mass loss from these important stars.

**Mass Loss in Binary Stars.** Two of the brightest and most famous stars in the sky exemplify both the far reaching issue of mass loss in binary systems and our continued poor understanding of the process. Sirius is a 2.3 $M_o$ A1V main sequence star with a 1.1 $M_o$ white dwarf companion in an orbital period of 50 years at a separation of 20.2 AU. Procyon is a similar case with a 1.74$M_o$ F5V primary in orbit with a 0.63 $M_o$ white dwarf and an orbital period of 40 years. The orbits of these two stars are wide enough that the present-day white dwarf in each evolved as a single star, thus it was originally the more massive member of the pair. Mass loss in these binaries occurred sometime in the astrophysical recent past expelling 2-3 and 1-2 $M_o$ of material respectively. Are the present day main sequence stars beneficiaries of the mass loss in that they were originally of lower mass? What are the true ages of these stars (Harrison et al. 2005, ApJ, 632, 123)?

An interesting case of fairly recent, large-scale mass loss occurs in the interacting binary stars termed Cataclysmic Variables. Here recent findings show that the present day lower mass stars (near 0.6-0.2 $M_o$) show CNO processed materials in their atmospheres arguing for a core evolution only possible in a far more massive star (Howell et al. 2001, ApJ, 550, 897). These same stars are too large and too luminous with respect to normal main sequence stars of similar mass (i.e., they are less massive than their radii suggest). A possible progenitor forerunner to such evolved binaries could be the Algol systems, in particular the sub-class called post-Algols. Algol binaries have long been held up as a prime example of a binary system which has undergone odd, unknown phases of mass loss. The "Algol Paradox", restated as "which present day star in the binary started out as the most massive", has confounded astronomers for over 50 years. Sudden orbital period changes of up to a few minutes and mass ratio values which appear unrealistic have been noted for Algols and ascribed to mass-loss events.

Interferometric measurements of binary systems, whereby the masses and other parameters can be established dynamically (not by photometric or spectroscopic means which are biased by gas streams, low gravity shells, and accretion disks), will make use of reflex motions to disentangle the true stellar characteristics. This is nicely illustrated in the case of β Lyr (Figure 4).

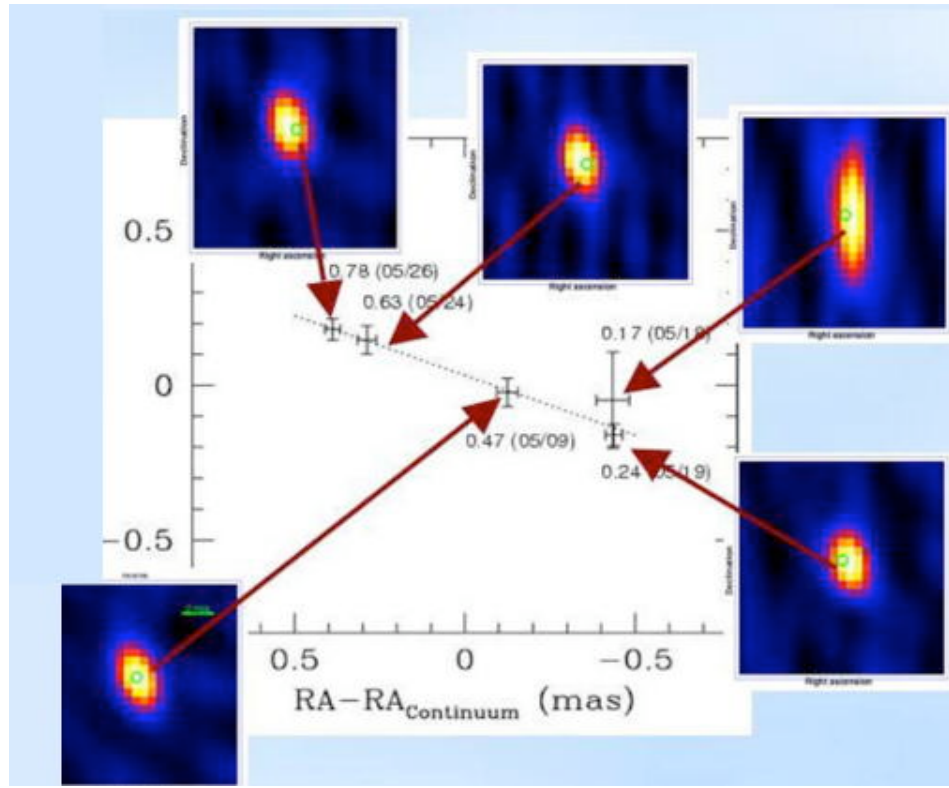

**Figure 4.** Reconstructed images of the β Lyr mass transfer system. These H-α images show the location of the HII emission region with respect to the star (position indicated by small circle) (Schmitt et al. 2009, ApJ 691, 984). These data reveal the orbit of the binary, which was hitherto misunderstood due to the obscuring mass-transfer stream.

Interferometry also promises rapid progress in the case of other systems in which companions are known to be important (novae, eg. pre-SN Type Ia systems such as RS Ophiuchi (Chesneau *et al.*, 2008, A&A 464, 119) or may have an important role, such as Be Stars, W-R stars, and planetary nebulae.

**The Future of Interferometric O/IR Imaging.** The Study of these and other types of mass-losing stars can extend the early studies discussed above down many avenues of higher spatial resolution (sub-milliarcsecond), improved imaging fidelity (tens of pixels instead of a few, dynamic ranges of a few hundred), wavelength coverage (optical through mid-IR), spectral resolution (greater than R ~10,000) and temporal exploration, with existing and developing instrumentation. Satisfactory modeling of mass loss has the potential to revolutionize our understanding of stars yet again, and will be needed to confidently apply stellar physics to stellar systems, and to trace nucleosynthetic products through galactic evolution.